\documentclass[a4paper,11pt]{article}
\usepackage{pos}
\usepackage{cleveref}
\usepackage{wrapfig}

\newlength{\bibitemsep}
\newlength{\bibparskip}
\let\oldthebibliography\thebibliography
\renewcommand\thebibliography[1]{%
  \oldthebibliography{#1}%
  \setlength{\parskip}{\bibitemsep}%
  \setlength{\itemsep}{\bibparskip}%
}

\newcommand{\taon}{$\tau$-lepton }
\newcommand{\taons}{$\tau$-leptons }
\newcommand{\Offline}{$\overline{\textrm{Off}}$\hspace{.05em}\protect\raisebox{.4ex}{$\protect\underline{\textrm{line}}$}}

\newcommand{\lsim}{\mathrel{\hbox{\rlap{\lower.75ex \hbox{$\sim$}} \kern-.3em \raise.4ex \hbox{$<$}}}}
\newcommand{\gsim}{\mathrel{\hbox{\rlap{\lower.75ex \hbox{$\sim$}} \kern-.3em \raise.4ex \hbox{$>$}}}}
\newcommand{\nupyprop}{{\texttt nuPyProp}}
\newcommand{\nuSpaceSim}{\texttt{\ensuremath{\nu}{SpaceSim}}}
\newcommand{\nuleptonsim}{{\texttt nuLeptonSim}}

\title{$\nu$SpaceSim: A Comprehensive Simulation Package for Modeling the Measurement of Cosmic Neutrinos using the Earth as the Neutrino Target and Space-based Detectors}
\ShortTitle{$\nu$SpaceSim: Cosmic Neutrinos and Space-based Detectors}

\author*[a]{Mary Hall Reno}
\author[a,b]{John F. Krizmanic}

\affiliation[a]{Department of Physics and Astronomy, University of Iowa\\
  Iowa City, Iowa 52242 USA}

\affiliation[b]{Astroparticle Physics Laboratory, NASA Goddard Space Flight Center\\
Greenbelt, Maryland 20771 USA}

\onbehalf{for the $\nu$SpaceSim Collaboration}

\emailAdd{mary-hall-reno@uiowa.edu}
\emailAdd{john.f.krizmanic@nasa.gov}

\abstract{$\nu$SpaceSim is a highly-efficient (e.g., fast) module-based, end-to-end simulation package that models the physical processes of cosmic neutrino interactions that leads to detectable signals for sub-orbital and space-based instruments. Starting with an input flux of neutrinos incident on a user-specified geometry in the Earth, the flux of Earth-emergent leptons are calculated followed by their subsequent extensive air showers (EAS). Next, the EAS optical Cherenkov and radio emission, signal attenuation to the detector, and the detector response are modeled to determine the sensitivity to both the diffuse cosmic neutrinos and transient neutrino sources. Using the Earth as a tau neutrino target and the atmosphere as the signal generator effectively forms a detector with a mega-gigaton mass. Furthermore, \taon decays and neutrino neutral-current interactions within the Earth (re)generates a flux of lower energy tau neutrinos that can also interact in the Earth thus enhancing the detection probability.  $\nu$SpaceSim provides a tool to both understand the data from recent experiments such as EUSO-SPB2 as well as design/understand the performance  the next generation of balloon- and space-based experiments, including POEMMA Balloon with Radio (PBR) and the Payload for Ultrahigh Energy Observations (PUEO). In this paper the $\nu$SpaceSim software, physics modeling, and the cosmic neutrino measurement capabilities of example sub-orbital and space-based experimental configurations are presented as well as status of planned modeling upgrades.}

\ConferenceLogo{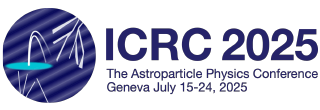}

\FullConference{39th International Cosmic Ray Conference (ICRC2025)\\
 15–24 July 2025\\
Geneva, Switzerland\\}

\begin{document}
\maketitle

\vspace{-5mm}
\section{Introduction}
\vspace{-3 mm}
\begin{figure}[!t]
    \centering
    \includegraphics[width=\textwidth]{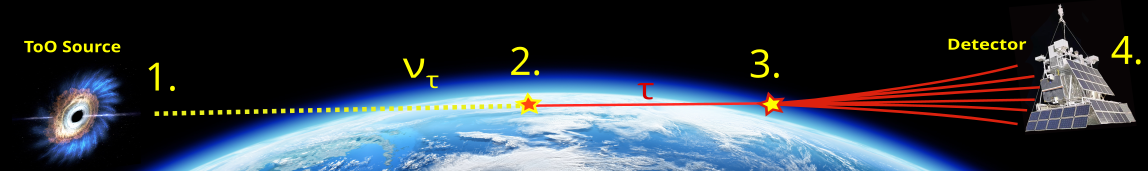}
    \caption{The Earth-skimming method for indirectly observing very-high-energy tau neutrinos (ToO observations pictured, fig. from ref. ~\cite{2024icrc.confE1134H}). 1-3: $\nu_{\tau}$ conversion to a $\tau$-lepton in the Earth, which emerges from the Earth. 3-4: $\tau$-lepton decays to EAS with signals that arrive at the detector.}
    \label{fig:ToO-method}
\end{figure}

$\nu$SpaceSim is a comprehensive end-to-end simulation software suite that models the detection of cosmic neutrinos by using the Earth as a neutrino target to produce Earth-emergent charged leptons. In the atmosphere, these produce extensive air showers (EAS) that generate optical and radio signals which can be measured by sub-orbital and space-based experiments. $\nu$SpaceSim has been successfully used by the science community to support the development of Ultra-Long-Duration-Balloon(ULDB)-borne experiments, such as EUSO-SPB2 \cite{2020NIMPA.95862164S,Eser:2023lck} and POEMMA Balloon with Radio (PBR) \cite{2024NIMPA106969819B} while also supporting the data analysis from the 2023 EUSO-SPB2 ULDB flight \cite{2024icrc.confE1134H} and ground-based observatories such as the Pierre Auger Observatory~\cite{barrera2025nuspacesim}. These include  modeling these experiments' responses to cosmic (anti-)neutrinos from diffuse fluxes or point sources including Target-of-Opportunity (ToO), observations illustrated in \cref{fig:ToO-method}.  

The $\nu$SpaceSim source code used in these analyses is publicly available via a web portal hosted by the HEASARC at GSFC \cite{HEASARC} and on github \cite{gitnuspacesim}.  $\nu$SpaceSim effectively models all aspects of the processes that start with a neutrino flux that produces neutrino interactions in the Earth through the Earth-emergent lepton-induced EAS signals to detection via an instrument response model after the signals are propagated through the atmosphere.  A sampled library approach and parallel (e.g., multi-core) processing efficiently generates the results for simulated signals at a specific location where the detector response module records the events. $\nu$SpaceSim is being developed in close coordination with the larger astroparticle physics community to produce the most useful and efficient tool-set in an adaptable framework to determine an instrument's neutrino sensitivity to both diffuse and point-source fluxes. Thus, current and future NASA experimental efforts in space-based cosmic neutrino detection, including EUSO-SPB2 \cite{2020NIMPA.95862164S}, PBR \cite{2024NIMPA106969819B}, the Probe Of Extreme Multi-Messenger Astrophysics (POEMMA) \cite{2021JCAP...06..007P}, the Antarctic Impulsive Transient Antenna (ANITA) \cite{2009APh....32...10A} and the Payload for Ultrahigh Energy Observations (PUEO) \cite{2020arXiv201002892A}, have benefited or will benefit from the results of $\nu$SpaceSim development. Work is underway to significantly boost the current version of the software by expanding functionality, implementing data-driven enhancements, and increasing the robustness by modeling systematic uncertainties within each of the $\nu$SpaceSim simulation modules. In this paper, some of these systematic uncertainties are quantified for a sample ULDB balloon instrument based on the optical Cherenkov design of EUSO-SPB2 telescope combined with geomagnetic EAS radio detection instrument.
\vspace{-5 mm}
\section{Configurations and Settings}
\vspace{-3 mm}

\begin{figure}[!t]
    \centering
    \includegraphics[width=\textwidth]{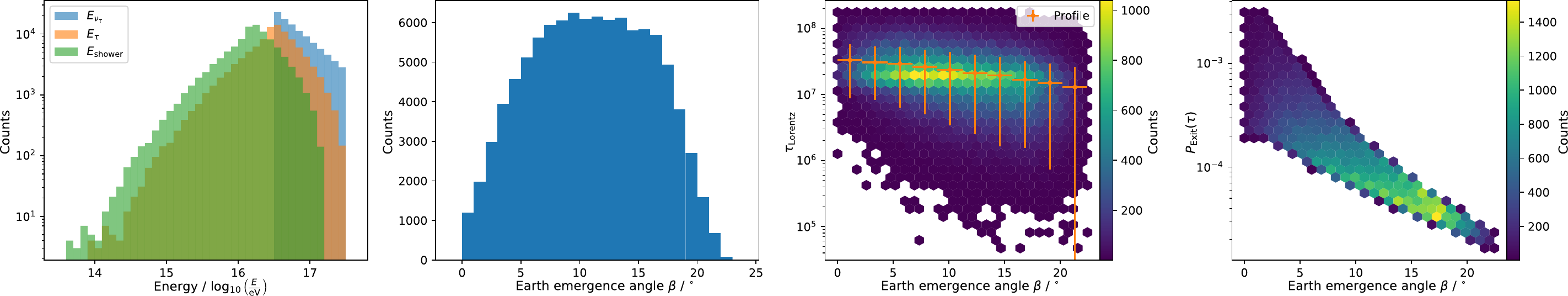}
    \caption{Sample analysis plots available in \nuSpaceSim\      (1) Energy distributions for an $E^{-2}$ spectrum of $\nu_{\tau}$s (blue) and resulting $\tau$-leptons (orange) and showers (green). (2)-(4) Distributions of $\tau$-lepton Earth-emergence angles, Lorentz factors, and exit probabilities.}
    \label{fig:dashboard}
\end{figure}

The \nuSpaceSim\ software is a command-line application that is based on a hierarchy of importable, high-performance Python modules. The user can create a TOML file (or use the default file) to set the detector altitude, response, thresholds, and geometric selection as well as modeling options such as the EAS development model used for the generation of optical or radio signals, Earth density model, and charged lepton interaction models. Intermediate and final data tables can be written in HEASARC's Flexible Image Transport System (FITS) and the Hierarchical Data Format (specifically HDF5). The input  baseline information provided by a user allows $\nu$SPaceSim to calculate the detector's acceptance to diffuse neutrinos or neutrino point sources for a given energy or a spectrum. If a spectrum of neutrino energies is chosen, the spectrum-weighted results will be reported as shown in Fig.~\ref{fig:dashboard}, which shows some example results of the plotting interface available in $\nu$SpaceSim for visualization for testing and analysis.

\vspace{-5 mm}
\section{Currently available features including new developments}
\vspace{-3 mm}

In this section we highlight the modeling performance of $\nu$SpaceSim by presenting results for a example experimental configuration: a ULDB experiment with an Optical Cherenkov telescope with similar performance to that in EUSO-SPB2 \cite{2020NIMPA.95862164S,Eser:2023lck} but with azimuth coverage of 360$^\circ$ while observing to 6.4$^\circ$ below the Earth's limb and a Cherenkov threshold of 200 photons/m$^2$ incident on the telescope. This ULDB example also includes a co-located radio detection instrument consisting of ten antennae operating in the 50 to 500 MHz band with 1.8 dB gain, a signal-to-noise (SNR) threshold of 5, and a model of ionospheric dispersion is used.  These examples highlight the the unique feature of $\nu$SpaceSim: its ability to simultaneously modeling the optical Cherenkov and geomagnetic signals from a common set of EAS.

\noindent{\bf Neutrino and \taon\ propagation} 
-- 
Currently \nuSpaceSim\ supports lepton-exit energy and probability look-up tables from \nupyprop\ \cite{nupyprop} and \nuleptonsim\ \cite{2025PhRvD.111b3012C}. 
Recent updates to the simulation program include models of the Earth’s density distribution \cite{DZIEWONSKI1981297,Kennett:1995ak}, and effects of tau depolarization \cite{Arguelles:2022bma} as \taons lose energy in the Earth~\cite{NuSpaceSim:2023ims}.

\begin{figure}[!t]
    \vspace{-2mm}
    \centering
   \includegraphics[width=0.55\linewidth]{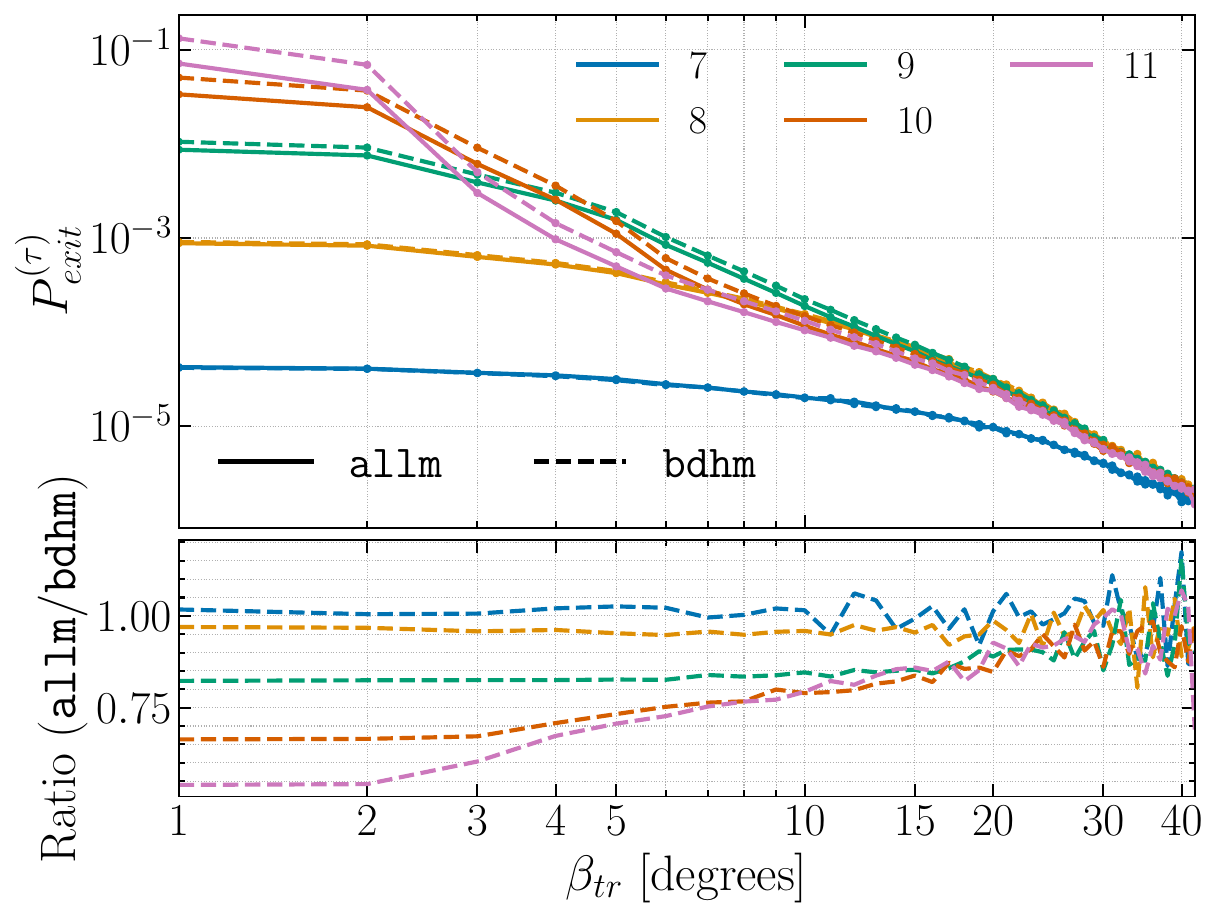}
    \includegraphics[width=0.44\linewidth]{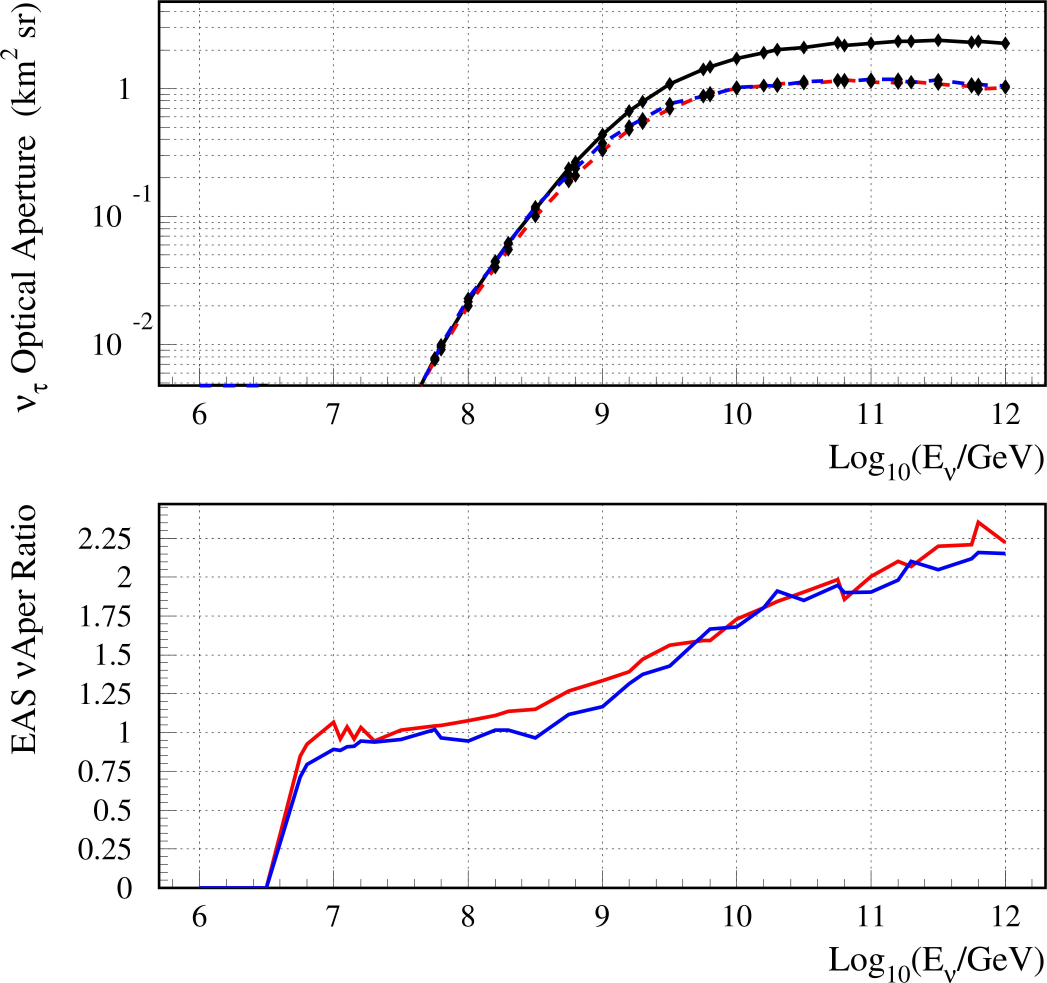} 
   \vspace{-5pt}

\vspace{-5mm}
    \caption{Left: The $\tau$-lepton exit probability for $\log_{10}(E_{\nu_\tau}/{\rm GeV})=7-11$ as a function of Earth emergence angle $\beta_{\rm tr}$ using the default electromagnetic energy loss model (ALLM) and alternate high energy extrapolation of energy loss (BDHM). Figure from ref. \cite{Garg:2022ugd}.
   Right: Diffuse neutrino flux optical (upper) $\nu_\tau$ aperture for the ULDB example instrument comparing the results for nuPyProp BHDM (black), nuPyProp ALLM (red) and nuLeptonSim (blue) as the neutrino to lepton generation models.The ratios (lower) of the BDHM/ALLM (red) and BDHM/nuLeptonSim (blue) quantifies how the BDHM model enhances the acceptance as the incident neutrino energy increases.} 
    \label{fig:energy-loss-models}
    \vspace{-5mm}
\end{figure}

A key feature of \nupyprop\ is that it is a standalone package. Its outputs are look-up tables (LUT) of lepton exit probabilities and lepton energy distributions, each as functions of incident neutrino energy and the lepton’s Earth emergence angle. The package has five high energy extrapolations of the neutrino nucleon cross section and two models of the Earth's density distribution as a function of radius. \nupyprop-1.0 LUTs are available for one neutrino cross section extrapolation and two charged lepton energy loss models, each for 42 individual Earth emergence angles and 25 incident neutrino energies that range for $E_\nu=10^6-10^{12}$\,GeV. These are generated with $10^8$ incident neutrinos per energy and angle. Fig.
\ref{fig:energy-loss-models} shows the impact of two  electromagnetic energy loss models  (ALLM and BDHM) on \taon exit probabilities, a component of the systematic uncertainty. The right panel of \cref{fig:energy-loss-models} shows the $\nu_\tau$ aperture for nuPyProp BDHM vs ALLM vs nuLeptonSim models of neutrino interactions and \taon\ electromagnetic energy loss for the ULDB example.

\begin{wrapfigure}{r}{0.5\textwidth}
    \begin{center}
    \vspace{-2mm}
    \includegraphics[width = 0.49\textwidth]{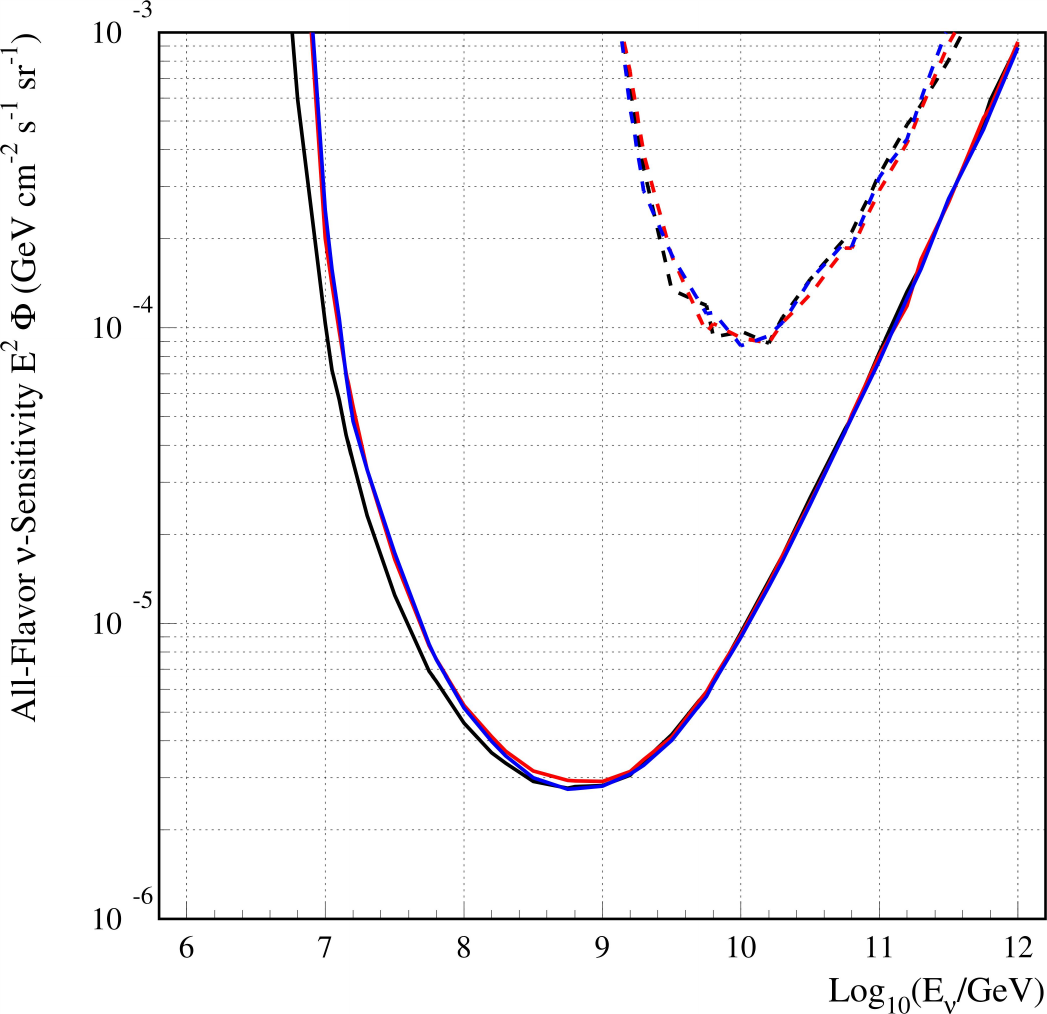}
    \vspace{-4mm}
    \caption{The all-flavor 90\% CL (per decade) diffuse flux sensitivity for the ULDB example instrument for the optical signal (solid) and radio (dashed) for three different EAS models: Greisen parameterization (black), Gaisser-Hillas Parameterization (red), and CONEX generated with EAS variability (blue).
    \vspace{-5ex}}
    \label{fig:EASlimComp}
    \end{center}
\end{wrapfigure}

\vspace{1mm}
\noindent{\bf EAS and Cherenkov Light Modeling} 

$\nu$SpaceSim currently supports three different EAS longitude profile models: the Greisen Parameterization 
\cite{Hillas2}, the Gaisser-Hillas Parameterization \cite{Gaisser-Hillas-1977-ICRC-15-8-353}, and CONEX \cite{CONEX} generated profiles that are provided via an external library. The comparison of the 90\%CL for the example ULDB instrument is shown in Fig.~\ref{fig:EASlimComp} and show the systematic variability is relatively small.  The baseline Cherenkov light model is one base on Hillas parameterizations \cite{Hillas1,Hillas2}.

\vspace{1mm}
\noindent{\bf Radio signal modeling} 
-- The radio signal modeling in \nuSpaceSim\ is based on ZHAireS  \cite{Alvarez-Muniz_2012} and RASPAS \cite{Tueros_2025} simulations. They model electromagnetic emission of particle tracks at the microscopic level within an EAS. A new method using a thinning procedure with tracks to build effective currents, which in turn are used to calculate the vector potential, has been developed for relatively fast and accurate calculations of radio emission \cite{Cummings:2023tuh}. These new simulations have been used to generate lookup tables of time-domain electric field voltages. Improvements are in progress to expand detector and shower configurations that can be simulated.

\begin{figure}
    \centering

       \includegraphics[width=0.8\textwidth]{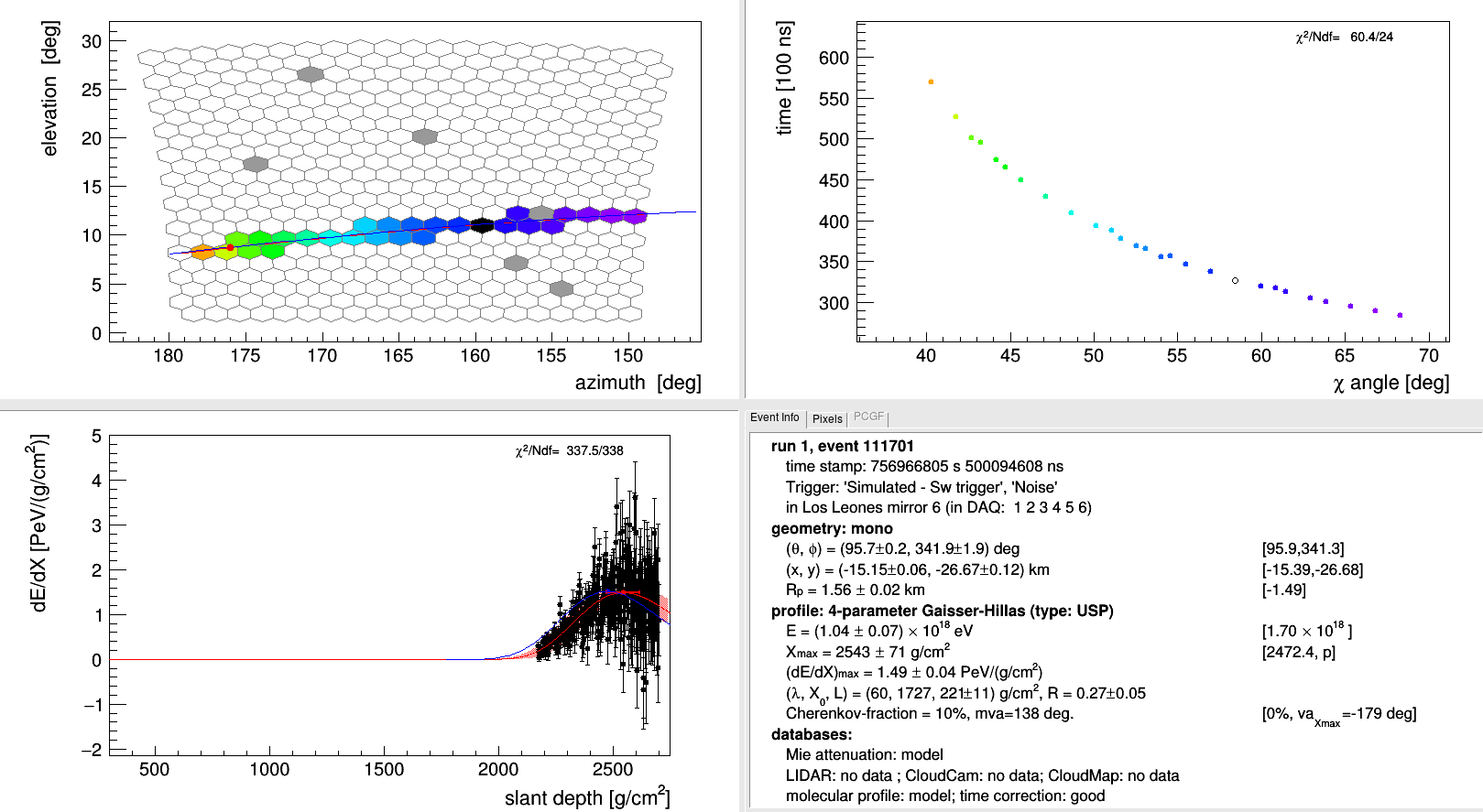}
 \captionof{figure}{ Simulated trigger of a \nuSpaceSim -generated shower in one of Pierre Auger's FD. The shower energy is 1.7$\cdot 10^{18}$ eV and its emergence angle is $5.9^\circ$ above the surface. Top left: Trace of the shower left in the Camera's pixels, showing a slightly upward-going trajectory. Top right: Timing distribution of triggering pixels. The color shows the time of each pixel's measurement (purple early, red late).
Bottom left: Energy deposition profile fit
Bottom right: Reconstruction results.}\label{fig:augertrigger} 
\end{figure}

\vspace{1mm}
\noindent{\bf Interface with external existing simulation framework} 
-- While \nuSpaceSim's output is provided by default as a FITS table, it can also be written to a ROOT file that emulates the structure of CONEX \cite{pierog2004first}. This functionality allows \nuSpaceSim\ to be integrated into a wide range of experiments whose simulation frameworks are already compatible with CONEX. As a prime example, this has enabled \nuSpaceSim\ to interface with the Pierre Auger Observatory's simulation framework, \Offline\ \cite{argiro2007offline}.  Figure \ref{fig:augertrigger} shows the simulation in \Offline\ of an upward-going shower generated by NuSpaceSim triggering one of the Observatory's Fluorescence detectors.

This framework was adopted and adapted for the EUSO Collaboration to support balloon-borne or space-based experiments like EUSO-SPB2, PBR, or POEMMA \cite{abe2024euso}. 
Current work is in progress in collaboration with the Pierre Auger Observatory to calculate the exposure of its FDs to neutrino-induced upward-going EAS. This work 
will complement a very recent work of the Auger collaboration following up on the ANITA upgoing EAS candidate events~\cite{PierreAuger:2025spg}.

\vspace{1mm}
\noindent{\bf Modeling of Sensitivity to Neutrino Point Sources} 

\begin{wrapfigure}{r}{0.5\textwidth}
    \begin{center}
   \vspace{-2mm}
    \includegraphics[width = 0.5\textwidth]{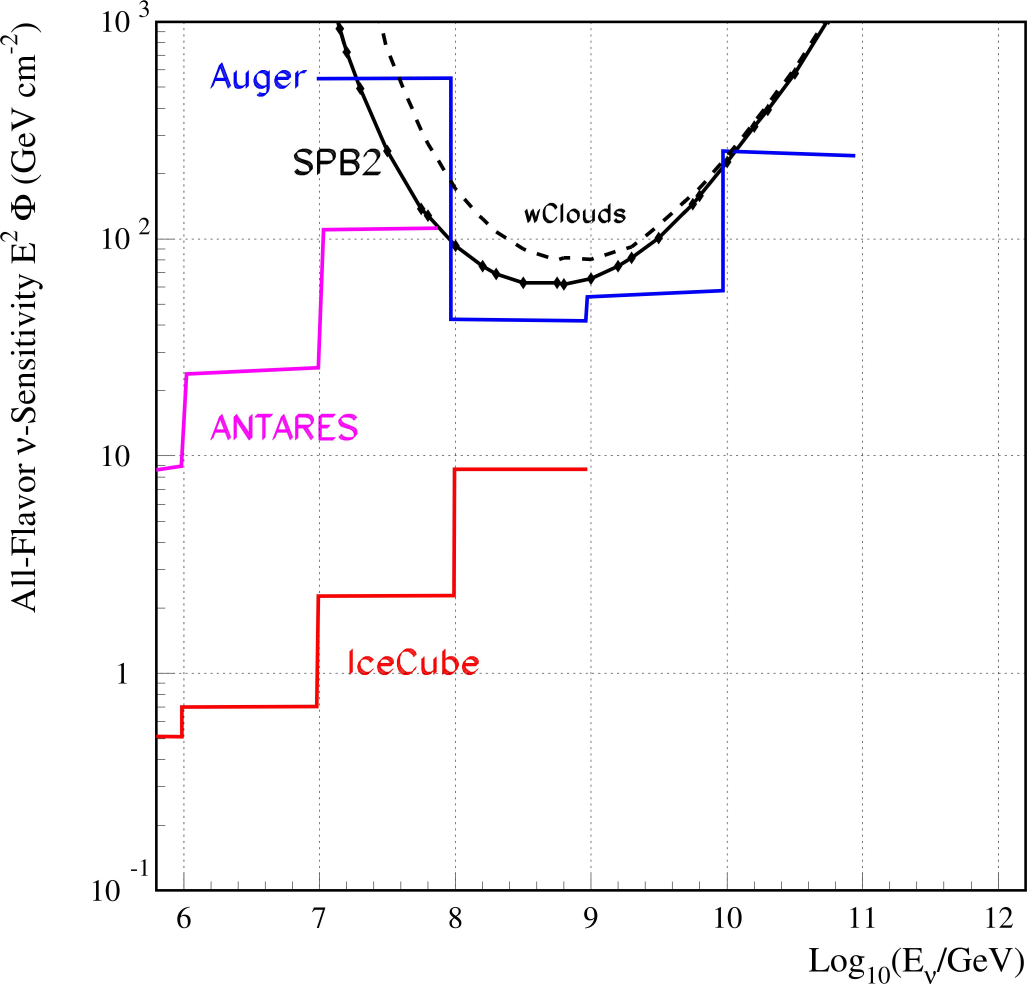}
    \caption{The all-flavor 90\% CL (per decade) for the ULDB example instrument for observing GW170817 compared the published limits \cite{2017ApJ...850L..35A} assuming 14 days of observation. The solid black labeled SPB2 is for sensitivity for the optical signal and  the dashed black curve is optical sensitivity assuming the average cloud cover in the instrument's field-of-view for August 2017 obtained from the MERRA-2 database.}
    \label{fig:GW170817}
    \end{center}
    \vspace{-8mm}
\end{wrapfigure}

$\nu$SpaceSim models an instrument sensitivity to either diffuse or astrophysical point sources, i.e., Target-of-Opportunity (ToO) events.
A User supplies the configuration parameters via the TOML input file that determine the calculation to be performed (diffuse or target-of-opportunity, ToO), the observational scenario and conditions (e.g., celestial coordinates of a source, observation time, cloud coverage), and detector characteristics (e.g., effective area, field of view, signal threshold) and trajectory parameters (e.g., time-stamped detector coordinates, altitude). Fig.~\ref{fig:GW170817} is an example of the modeling of a ToO observation, the ULDB instrument was positioned over Wanaka, New Zealand and the optical sensitivity to observing GW170187 was modeled assuming a 14-day observation, with and without the effect of clouds.

\vspace{-3 mm}

The ULDB example used in these calculations are based on the EUSO-SPB2 and PBR Cherenkov telescopes but with a larger, 360$^\circ$ azimuth viewing range, compared to the 12.8$^\circ$ for EUSO-SPB2 and similar for PBR.
Fig. \ref{fig:GW170817} shows the importance of modeling observing conditions for understanding sensitivity limits of neutrino astrophysical transient events. 
$\nu$SpaceSim will be a powerful tool for predicting and understanding PBR's neutrino measurement performance.
These results plus the inclusion of geomagnetic radio techniques highlight $\nu$SpaceSim's essential role in optimizing instrument designs and observation campaigns including those that combine optical and radio detection.

\vspace{-5 mm}
\section{Developments in Progress}
\vspace{-3 mm}

$\nu$SpaceSim work to support the EUSO-SPB2 data analysis has provided focus on near-term improvements.  These include modeling a detector's channel-by-channel variability, the finer details of the optical Cherenkov light variability,  more robust modeling of the EAS variability, the viewing geometry to model the observation of above-the-limb EAS induced by cosmic rays, include motion of the payload, completing the radio ToO module, and developing a model for a combined optical-radio trigger.  Some specifics follow.

\vspace{1mm}
\noindent {\bf Cherenkov light modeling and interface }-- 
The model for the generation of Cherenkov light is currently being upgraded to use CHASM \cite{buckland2023universality}. CHASM is a simulation package that exploits the universality of charged particles in EAS to calculate the Cherenkov photon yield and angular distribution by sampling the EAS profile throughout its many stages and altitudes. The Cherenkov photon production is specifically calculated for specified detector positions with respect to the EAS location and direction. CHASM simulations agree with the signal generated by CORSIKA-IACT \cite{CORSIKA}, but uses universality-based look up tables rather than an explicit particle stack to run in fractions of a second over all EAS energies. CHASM approximates the wavelength-dependent atmospheric attenuation of Cherenkov photons in a curved atmosphere using the CORSIKA-IACT Cherenkov extinction table. CHASM can write out the temporal signal of Cherenkov photons to the CORSIKA-IACT EventIO, to AstroPy tables \cite{astropy2022}, or to its own Python format. These output formats will allow for additional interfacing of \nuSpaceSim\  with the simulation frameworks of experiments with Cherenkov detectors, such as PBR and POEMMA, supported by EUSO \Offline. 

\vspace{1mm}
\noindent
{\bf Modeling atmospheric properties} --
The detectors simulated in \nuSpaceSim\ use the atmosphere as the sensitive volume, so it is important to model properties of the atmosphere. Aerosols and clouds are especially relevant for optical signal propagation. The radio signal remains largely unaffected by atmospheric conditions. We currently use MERRA-2 database \cite{MERRA2} to determine the average cloud coverage for any point on the globe and will extend the use of MERRA-2 to include aerosols and ozone.

\vspace{1mm}
\noindent{\bf Balloon/Satellite motion} The current version of \nuSpaceSim{} does not include the motion of a balloon or satellite detector. However, an interface with the Neutrino-Target-Schedulers software is being developed, which will provide data of the detector location and detector motion has on the relative location of ToO sources in the sky on real or simulated flight paths. 

\vspace{-5 mm}
\section{Summary and acknowledgment}
\vspace{-3 mm}

The central goal of the \nuSpaceSim\ software package is to provide end-to-end simulation results that the cosmic-ray and neutrino communities can use to design suborbital and space-based experiments, plan observations, and assist in the analysis of the data. Fittingly, \nuSpaceSim\ has already become a unique resource for efficient simulating upward-moving EASs within collaborations such as EUSO-SPB2, POEMMA, and PBR. As described, enhancements to the existing software are underway, with the goal of expanding the user base which will provide feedback for $\nu$SpaceSim improvements.

This work is supported by NASA RTOP 21-APRA21-0071 at
NASA/GSFC and NASA grants 80NSSC22K1520 at the University of Chicago, 80NSSC22K1517
at the Colorado School of Mines, 80NSSC22K1523 at the University of Iowa, 80NSSC22K1519 at
Pennsylvania State University, and 80NSSC22K1522 at the University of Utah.

\vspace{-5mm}


\clearpage
\section*{$\nu$SpaceSim Collaboration}
%
%
%

\scriptsize
\noindent
John  Krizmanic.$^1$,
Yosui Akaike$^2$, 
Luis Anchordoqui$^3$, 
Douglas Bergman$^4$, 
Isaac Buckland$^4$, 
Jorge Caraca-Valente$^5$,
Austin Cummings$^6$,
Johannes Eser$^7$,
Fred Angelo Batan Garcia$^7$,
Diksha Garg$^9$,
Claire Gu\'epin$^{10}$,
Tobias Heibges$^5$,
Luke Kupari$^9$,
Andrew Ludwig$^{11}$,
Simon Mackovjak$^{12}$,
Eric Mayotte$^5$,
Sonja Mayotte$^5$,
Angela Olinto$^7$,
Thomas Paul$^7$, 
Alex Reustle$^{1,13}$,
Andrew Romero-Wolf$^{11}$,
Mary Hall Reno$^9$,
Fred Sarazin$^{5}$,
Tonia Venters$^{1}$
Lawrence Wiencke$^{5}$,
Stephanie Wissel$^6$ \\

%
\noindent

$^{1}$NASA/Goddard Space Flight Center, Greenbelt, Maryland 20771 USA,
$^2$Waseda Institute for Science and Engineering, Waseda University, Shinjuku, Tokyo, Japan,
$^3$Department of Physics and Astronomy, Lehman College, City University of New York, New York, New York, 10468 USA,
$^4$Department of Physics and Astronomy, University of Utah, Salt Lake City, Utah 84112 USA,
$^{5}$Department of Physics, Colorado School of Mines, Golden, Colorado 80401 USA
$^6$Department of Physics, Pennsylvania State University, State College, Pennsylvania 16801 USA,
$^7$Department of Astronomy and Astrophysics Columbia University, New York, New York 10027 USA,
$^8$Department of Physics and Astronomy, University of Iowa, Iowa City, Iowa 52242 USA,
$^9$Department of Astronomy, University of Maryland, College Park, College Park, Maryland 20742 USA,
$^{10}$Laboratoire Univers et Particules de Montpellier (LUPM), Universit\'e Montpellier,
Montpellier Cedex 5, France,
$^{11}$Jet Propulsion Laboratory, California Institute of Technology, Pasadena, California 91109, USA,
$^{12}$Institute of Experimental Physics, Slovak Academy of Sciences, Kosice, Slovakia,
$^{13}$ADNET Systems.

\end{document}